\documentclass{ws-ijmpb}
\usepackage{amsfonts,amssymb,amsmath}
\usepackage{graphicx,epstopdf}
\usepackage[super,sort,compress]{cite}

\begin{document}

\markboth{BING WANG}
{THE TRANSPORT OF SELF-PROPELLED ELLIPSOIDAL PARTICLES CONFINED IN $2D$ SMOOTH CORRUGATED CHANNEL}

%
\catchline{}{}{}{}{}
%

\title{THE TRANSPORT OF SELF-PROPELLED ELLIPSOIDAL PARTICLES CONFINED IN $2D$ SMOOTH CORRUGATED CHANNEL }

\author{BING WANG}

\address{School of Mechanics and Optoelectronic Physics, Anhui University of Science and Technology \\ Huainan, 232001, P.R.China\\
hnitwb@163.com}

\author{WENFEI WU}
\address{School of Mechanics and Optoelectronic Physics, Anhui University of Science and Technology \\ Huainan, 232001, P.R.China}

\maketitle

\begin{history}
\received{Day Month Year}
\revised{Day Month Year}
\end{history}

\begin{abstract}
Directed transport of self-propelled ellipsoidal particles confined in a smooth corrugated channel with asymmetric potential and Gaussian colored noise is investigated. Effects of the channel, potential and colored noise on the system are discussed. Large $x$ axis noise intensity inhibits the transport in $-x$ and $+x$ direction. The directed transport speed $|\langle V\rangle|$ has a maximum with increasing $y$ axis noise intensity. Proper size of the bottleneck is good for the directed transport of the ellipsoidal particles, but large and small size of bottleneck inhibits this directed transport. The transport reverse phenomenon appears with increasing load and self-propelled speed. Perfect sphere particle is easier to directed transport than needlelike ellipsoid particle.
\end{abstract}


\section{Introduction}
Recent years, we have seen enormous activities in the study of the properties of the confined Brownian particles. These studies, both theoretical and experimental, revealed that there are two ways of confinement fundamentally effect a system. One way is by regulating the transport space accessible to its diffusing components\cite{PHanggi}, and the other way is by increasing the hydrodynamic drag on the particles\cite{Burada}. There exists a large number of natural and artificial confined geometries, e.g., biological cells\cite{HXZhou,YWu,JLiu,Cui2020}, zeolites\cite{Keil}, artificial nanopores\cite{Firnkes,Pedone}, ionpumps\cite{Siwy,ZSiwy,Bosi,Ghosh} and micro fluidic devices\cite{Matthias,Squires,Bradley,Dagdug}.

Confined particle shows a series of novel features, e.g. current reversal\cite{Jia,Hu,Liu,Xu,HWHu}, self-organization\cite{Richardi,Yoshida}, geometry-induced stochastic resonance phenomenon\cite{Ghosh020601,Ghosh011109} and so on. H\"{a}nggi \emph{et al}. proposed a model of asymmetry particles confined in a compartmentalized channel and found the absolute negative mobility\cite{Marchesoni}. Ghosh \emph{et al}. investigated directed transport of suspended particles and found inertial corrections must be considered when the width of the bottlenecks is smaller than an appropriate particle diffusion length\cite{Ghosh2012}. Using the hybrid molecular dynamics method, Chen \emph{et al}. studied the properties of self-propelled synthetic motors\cite{Chen2018}. Pu \emph{et al}. investigated the reentrant phase separation behavior of active particles and found that phase separation shows a re-entrance behavior with variation of the interaction strength\cite{Pu}. Li \emph{et al}. found non-Gaussian normal diffusion phenomenon of Brownian particles floating in a narrow corrugated channel with fluctuating cross section\cite{Li2019}. Yang \emph{et al}. experimentally investigated the diffusion of particles moving in a planar channel\cite{Yang2019}.

In many biological systems, transport particles are often nonspherical, such as proteins diffusing in membranes and fine grains migrating through the pores of micro media\cite{Lugli,Saffman}. Han \emph{et al}. studied the transport of ellipsoid particles confined in a two dimensional channel and quantified the crossover from short-time anisotropic to long-time isotropic diffusion\cite{Han}. Ohta \emph{et al}. found deformable particles exhibit bifurcation and circular motion phenomenon\cite{Ohta}. Ai \emph{et al}. investigated the rectified transport of active ellipsoidal particles in a two-dimensional asymmetric potential\cite{Ai}. Ghosh \emph{et al}. numerically simulated the transport of elliptic particles confined in two-dimensional channels with reflecting walls and observed long diffusion transients\cite{Ghosh062115}. Traditionally, stochastic differential equations used in physical and biological science have involved Gaussian white noise. Investigation of laser noise problems \cite{Zhu,XZhou}, bistable systems\cite{Billah} and self-propelled particle systems\cite{Wittmann,Mondal} found that it is necessary to consider colored noise in those systems.

In this paper, we investigate the directed transport of self-propelled ellipsoidal Brownian particles confined in a two dimensional($2D$) smooth channel. The paper is organized as follows: The basic model of the system is provided in Sect.\ref{label2}. In Sect.\ref{label3}, the effects of the channel, the potential and the colored noise on the ellipsoidal particles are investigated by means of simulations. In Sect. \ref{label4}, we get the conclusions.

\section{\label{label2}Basic model and methods}
In this work, we consider self-propelled ellipsoidal particles confined in a $2D$ smooth corrugated channel with potential and Gaussian colored noise. In the lab frame, the displacement $\delta \vec{R}(t)$ of the particle can be described by the mass center($\delta{x}$, $\delta{y}$). We use the following Langevin equations to describe the dynamics of the particle\cite{Han,Grima}

\begin{equation}
\frac{\partial {x}}{\partial t}=v_0\cos\theta(t)+F_x[\bar{\Gamma}+\Delta\Gamma\cos 2\theta(t)]+\Delta\Gamma F_y\sin2\theta(t)+\xi_x(t), \label{Ext}
\end{equation}

\begin{equation}
\frac{\partial {y}}{\partial t}=v_0\sin\theta(t)+F_y[\bar{\Gamma}-\Delta\Gamma\cos 2\theta(t)]+\Delta\Gamma F_x\sin2\theta(t)+\xi_y(t), \label{Eyt}
\end{equation}

\begin{equation}
\frac{\partial {\theta(t)}}{\partial t}={\xi}_\theta(t). \label{Ethetat}
\end{equation}
The angle between the lab frame $x$ axis and the body frame $\hat{x}$ axis is $\theta(t)$. The self-propelled velocity is $v_0$, and $v_0$ is along the long axis of the particle. The quantities $\bar{\Gamma}=\frac{1}{2}(\Gamma_x+\Gamma_y)$ and $\Delta{\Gamma}=\frac{1}{2}(\Gamma_x-\Gamma_y)$ are the average and difference mobilities of the body, respectively. The mobilities along its long axis and short axis are $\Gamma_x$ and $\Gamma_y$, respectively. $\Delta\Gamma$ determines the asymmetry of the body of the ellipsoidal particle. The particle is a perfect sphere when $\Delta\Gamma=0$ and a very needlelike ellipsoid when $\Delta \Gamma\rightarrow\bar\Gamma$. $\xi_{x}$, $\xi_{y}$ and $\xi_{\theta}$ are the noises. $\xi_x$ and $\xi_y$ parallel to $x$ axis and $y$ axis, respectively. $\xi_{\theta}$ is the angle noise. $\xi_{x}$, $\xi_{y}$ and $\xi_{\theta}$ satisfy the following relations,
\begin{equation}
\langle\xi_i(t)\rangle=0,(i=x,y,\theta),
\end{equation}
\begin{equation}
\langle\xi_i(t)\xi_j(t')\rangle=\delta_{ij}\frac{Q_i}{\tau_i}\exp[-\frac{|t-t'|}{\tau_i}],(i=x,y,\theta),
\end{equation}
where $Q_i$ and $\tau_i$ are the noise intensity and the self-correlation time of the noises, respectively.

\begin{figure}
\center{
\includegraphics[height=8cm,width=10cm]{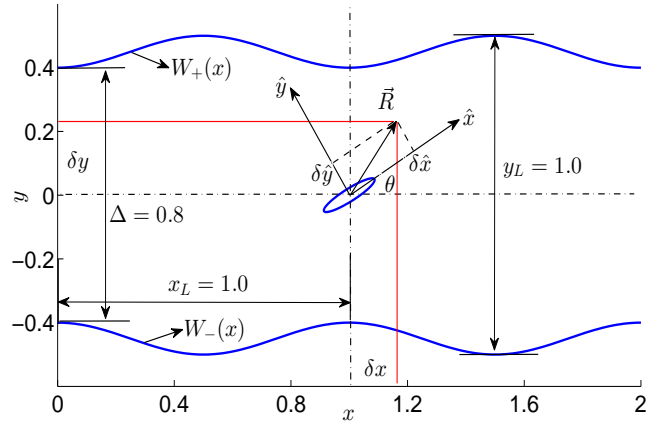}
\caption{Illustrations of the smooth corrugated channel with $\Delta=0.8$, $x_L=y_L=1.0$, $\eta=2.0$.}
\label{Channel}}
\end{figure}

The particles are confined in a two dimensional smooth corrugated channel. As shown in Fig.\ref{Channel}, the channel is consisted of many cavities and is periodic in spatial along the $x$-axis. The walls of the cavity are modeled by the following sinusoidal functions,\cite{Ghosh061162}
\begin{equation}
W_+(x)=\frac{1}{2} [\Delta +(y_L-\Delta)\sin^{\eta}(\frac{\pi x}{x_L})],
\end{equation}
\begin{equation}
W_-(x)=-\frac{1}{2} [\Delta +(y_L-\Delta)\sin^{\eta}(\frac{\pi x}{x_L})],
\end{equation}
where $x_L=1.0$ and $y_L=1.0$ are the length and width of the cavity, respectively. The additional tunable geometric parameter is $\eta$. The cavity represents the compartment of sinusoidally corrugated channel when $\eta=2$. When $\eta\rightarrow0$, the cavity reproduces the compartment of sharply corrugated channels. The channel width is $h(x)=W_+-W_-$. The minimal channel width(the bottleneck) is $h_{min}=h(x)|_{x=\pm k,k=0, 1, 2,\cdot\cdot\cdot}=\Delta$, and through which the particles can exit the cavity. The maximal channel width is $h_{max}=h(x)|_{x=\pm (2k+1)\frac{1}{2},k=0, 1, 2,\cdot\cdot\cdot}=y_L$.

\begin{figure}
\center{
\includegraphics[height=8cm,width=12cm]{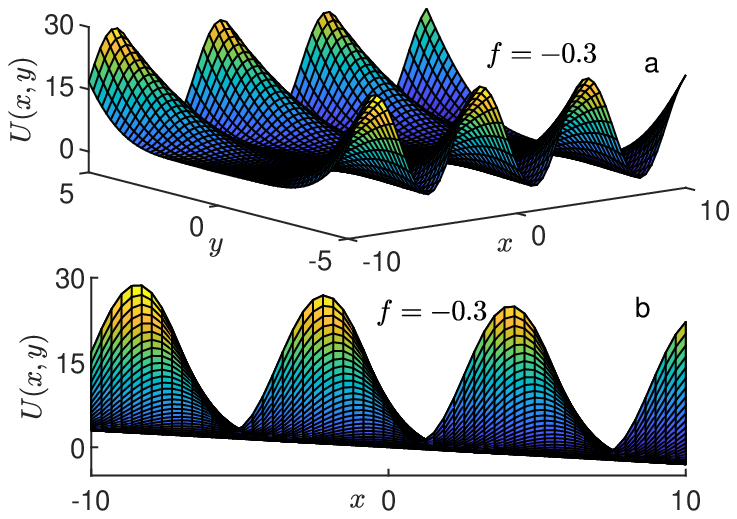}
\includegraphics[height=8cm,width=12cm]{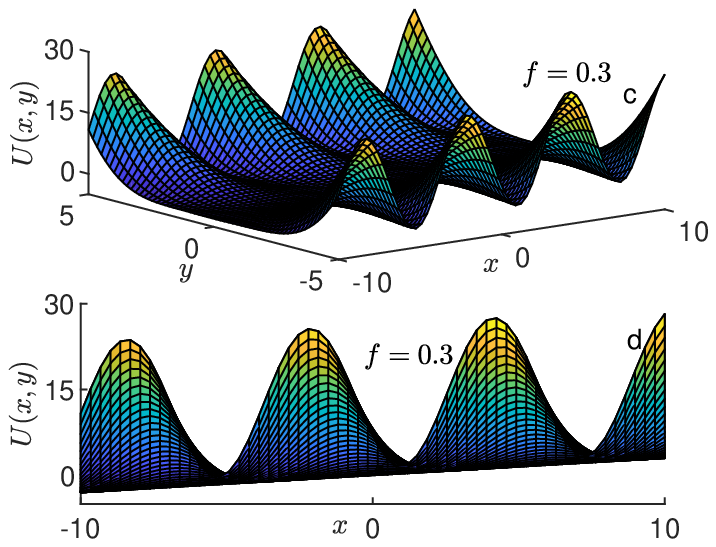}
\caption{The asymmetric potential $U(x,y)$ with $U_0=1.0$, $\varepsilon=0.5$:(a)3D view with $f=-0.3$;(b)Front view with $f=-0.3$; (c)3D view with $f=0.3$; (d)Front view with $f=0.3$.}
\label{Uxy}}
\end{figure}

The asymmetric potential is described by the following equation\cite{Ai}(shown in Fig.\ref{Uxy}),
\begin{equation}
U(x,y)=\frac{U_0}{2}y^2[\cos(x+\varepsilon \ln {\cosh y})+1.1]+fx, \label{Uxy}
\end{equation}
where $U_0$ is the height of the potential, and $f$ is the load. The asymmetric parameter of the potential is $\varepsilon$, and the potential is symmetric for $\varepsilon=0.0$. The equipotential is look like a herringbone pattern if $\varepsilon\neq0$. The forces $F_x=-\frac{\partial U}{\partial x}$ and $F_y=-\frac{\partial U}{\partial y}$ are along $x$ and $y$ directions of the lab frame, respectively.

In the theory of Brownian motion, the central question is the particle's over all long time behavior. The velocity of the particle is one of the key quantities. We only calculate the average velocity in $x$ direction because the channel in $y$ direction is confined,
\begin{equation}
\langle V_{\theta_0}\rangle=\lim_{t\to\infty}\frac{\langle{x(t)-x(t_0)}\rangle}{t-t_0},
\end{equation}
the position of particles at time $t_0$ is $x(t_0)$. The initial angle of the trajectory is $\theta_0$. The full average velocity after another average over all $\theta_0$ is
\begin{equation}
\langle V\rangle=\frac{1}{2\pi}\int^{2\pi}_0 \langle V_{\theta_0}\rangle d\theta_0.
\end{equation}
\section{\label{label3}Results and discussion}
In order to give a simple and clear analysis of the system. We integrate Eqs.(\ref{Ext},\ref{Eyt},\ref{Ethetat}) with time step $\Delta t=10^{-4}$ using the Euler algorithm. The average velocity is obtained as ensemble averages over $10^5$ trajectories with random initial conditions. In the simulation, we set $x_L=y_L=1.0$, $\bar\Gamma=1.0$ and $U_0=1.0$ throughout the paper.

\begin{figure}
\center{
\includegraphics[height=8cm,width=10cm]{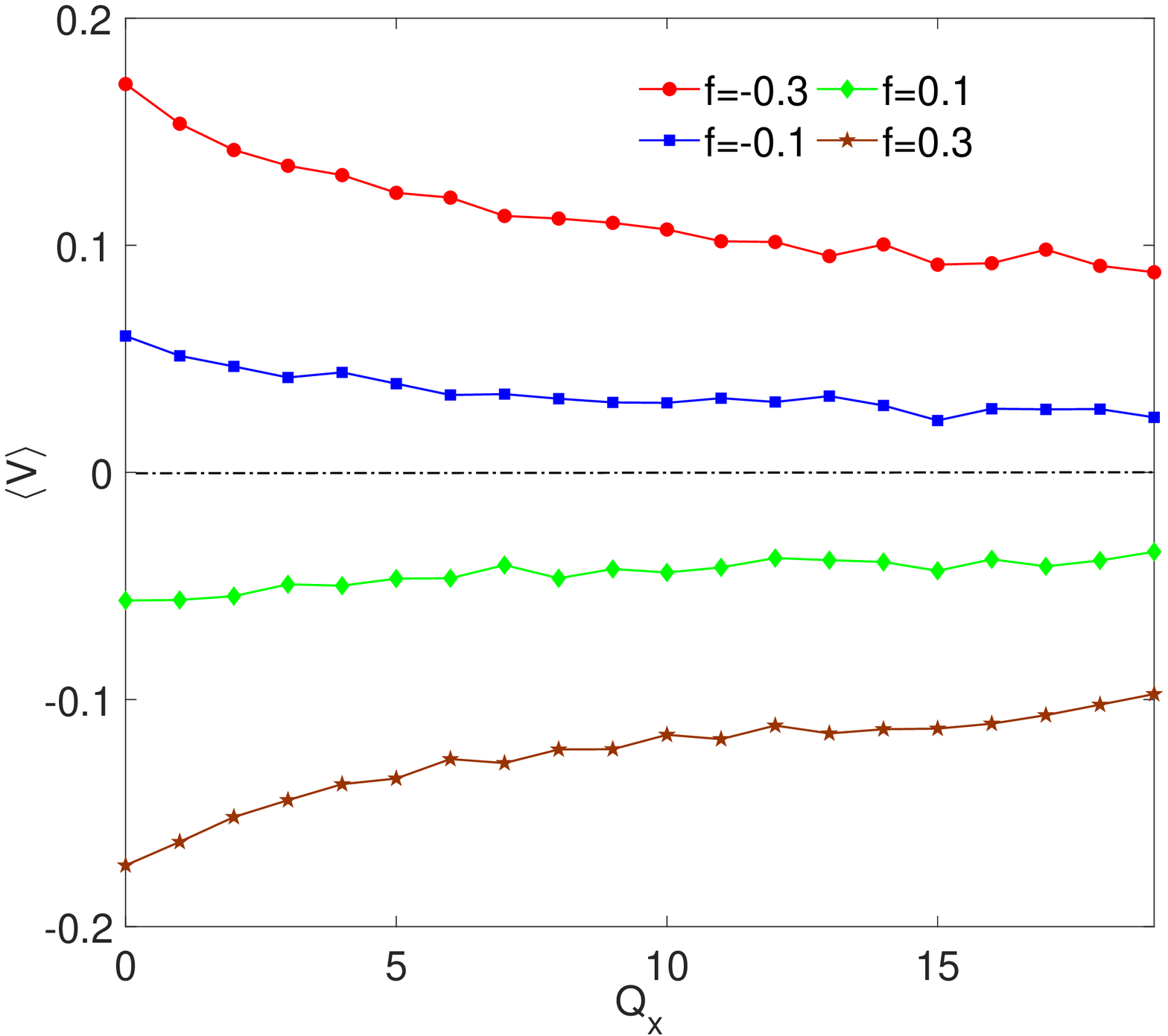}
\caption{The average velocity $\langle V\rangle$ as a function of $Q_x$ for different values of $f$. The other parameters are $\Delta=0.8$, $\varepsilon=0.5$, $\eta=2.0$, $v_0=0.5$, $Q_x=Q_y=Q_\theta=1.0$, $\tau_x=\tau_y=\tau_\theta=1.0$.}
\label{VQx}}
\end{figure}
Fig.\ref{VQx} displays the average velocity $\langle V\rangle$ as a function of $x$ axis noise intensity $Q_x$ with different load $f$. In this paper, we find the average velocity $\langle V\rangle>0$, and $\langle V\rangle$ decreases with increasing $Q_x$ when the load $f=-0.1$ and $f=-0.3$. Which means that negative load $f$ induces directed transport in $+x$ direction, and large $x$ axis noise intensity $Q_x$ inhibits this transport. When the load $f>0$($f=0.1$ and $f=0.3$), the average velocity $\langle V\rangle<0$, and $\langle V\rangle$ increases with increasing $Q_x$. So positive $f$ leads to directed transport in $-x$ direction, and large $Q_x$ is bad for this directed transport. Of cause, we can also say that the directed transport speed $|\langle V\rangle|$($|\langle V\rangle|$ is the absolute value of $\langle V\rangle$) decreases with increasing $Q_x$, so large $x$ axis noise intensity inhabits directed transport along $x$ axis.

\begin{figure}
\center{
\includegraphics[height=8cm,width=10cm]{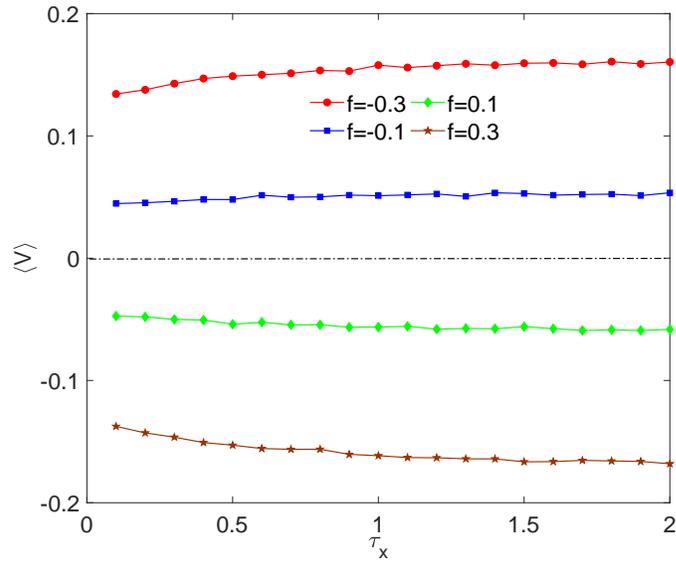}
\caption{The average velocity $\langle V\rangle$ as a function of $\tau_x$ for different values of $f$. The other parameters are $\Delta=0.8$, $\varepsilon=0.5$, $\eta=2.0$, $v_0=0.5$, $Q_x=Q_y=Q_\theta=1.0$, $\tau_y=\tau_\theta=1.0$.}
\label{VTaux}}
\end{figure}
Fig.\ref{VTaux} displays $\langle V\rangle$ as a function of self-correlation time $\tau_x$ with different load $f$. In this figure, comparing with the effect of $Q_x$ in Fig.\ref{VQx}, we find $\tau_x$ has exactly reverse effect on the system. The directed transport speed $|\langle V\rangle|$ increases with increasing $\tau_x$. Large self-correlation time $\tau_x$ is good for the transport. In this figure, we can also find negative $f$ induces directed transport in $+x$ direction, but positive $f$ induces directed transport in $-x$ direction. From Figs. \ref{VQx} and \ref{VTaux}, we find the slopes of $\langle V\rangle-Q_x$ and $\langle V\rangle-\tau_x$ curves almost changes to zero when $Q_x$ and $\tau_x$ are large. So changes of $Q_x$ and $\tau_x$ have weak impact on the transport when their values are large.

\begin{figure}
\center{
\includegraphics[height=8cm,width=10cm]{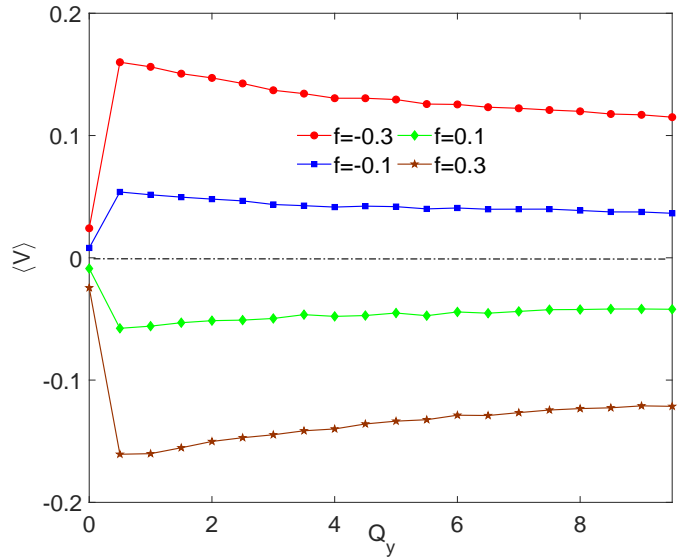}
\caption{The average velocity $\langle V\rangle$ as a function of $Q_y$ for different values of $f$. The other parameters are $\Delta=0.8$, $\varepsilon=0.5$, $\eta=2.0$, $v_0=0.5$, $Q_x=Q_\theta=1.0$, $\tau_x=\tau_y=\tau_\theta=1.0$.}
\label{VQy}}
\end{figure}
Fig.\ref{VQy} displays $\langle V\rangle$ as a function of $y$ axis noise intensity $Q_y$ with different $f$. In this figure, we find the directed transport speed $|\langle V\rangle|$ has a maximum with increasing $Q_y$. So proper $y$ axis noise intensity is good for the transport in $x$ and $-x$ direction, but too large or too small $Q_y$ inhabits this phenomenon. From figures \ref{VQx} and \ref{VQy}, we find an interesting phenomenon, large $x$ axis noise intensity inhabits the directed transport in $+x$ and $-x$ direction, but proper $y$ axis noise intensity is good for this transport.

\begin{figure}
\center{
\includegraphics[height=8cm,width=10cm]{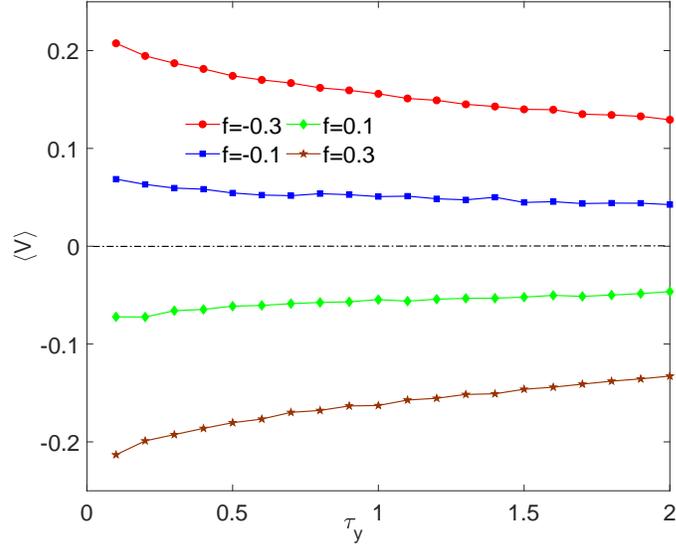}
\caption{The average velocity $\langle V\rangle$ as a function of $\tau_y$ for different values of $f$. The other parameters are $\Delta=0.8$, $\varepsilon=0.5$, $\eta=2.0$, $v_0=0.5$, $Q_x=Q_y=Q_\theta=1.0$, $\tau_x=\tau_\theta=1.0$.}
\label{VTauy}}
\end{figure}
Fig.\ref{VTauy} displays $\langle V\rangle$ as a function of $y$ axis noise self-correlation time $\tau_y$ for different $f$. Comparing with the effect of $\tau_x$ in Fig.\ref{VTaux}, we find the directed transport speed $|\langle V\rangle|$ decreases with increasing $\tau_y$. This means that large $\tau_y$ has negative effect on the transport along $x$ axis.

\begin{figure}
\center{
\includegraphics[height=8cm,width=10cm]{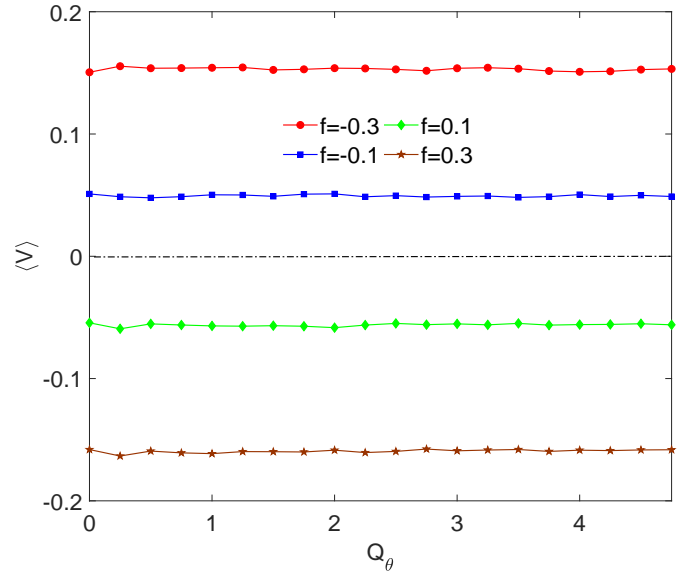}
\caption{The average velocity $\langle V\rangle$ as a function of $Q_\theta$ for different values of $f$. The other parameters are $\Delta=0.8$, $\varepsilon=0.5$, $\eta=2.0$, $v_0=0.5$, $Q_x=Q_y=1.0$, $\tau_x=\tau_y=\tau_\theta=1.0$.}
\label{VQtheta}}
\end{figure}

\begin{figure}
\center{
\includegraphics[height=8cm,width=10cm]{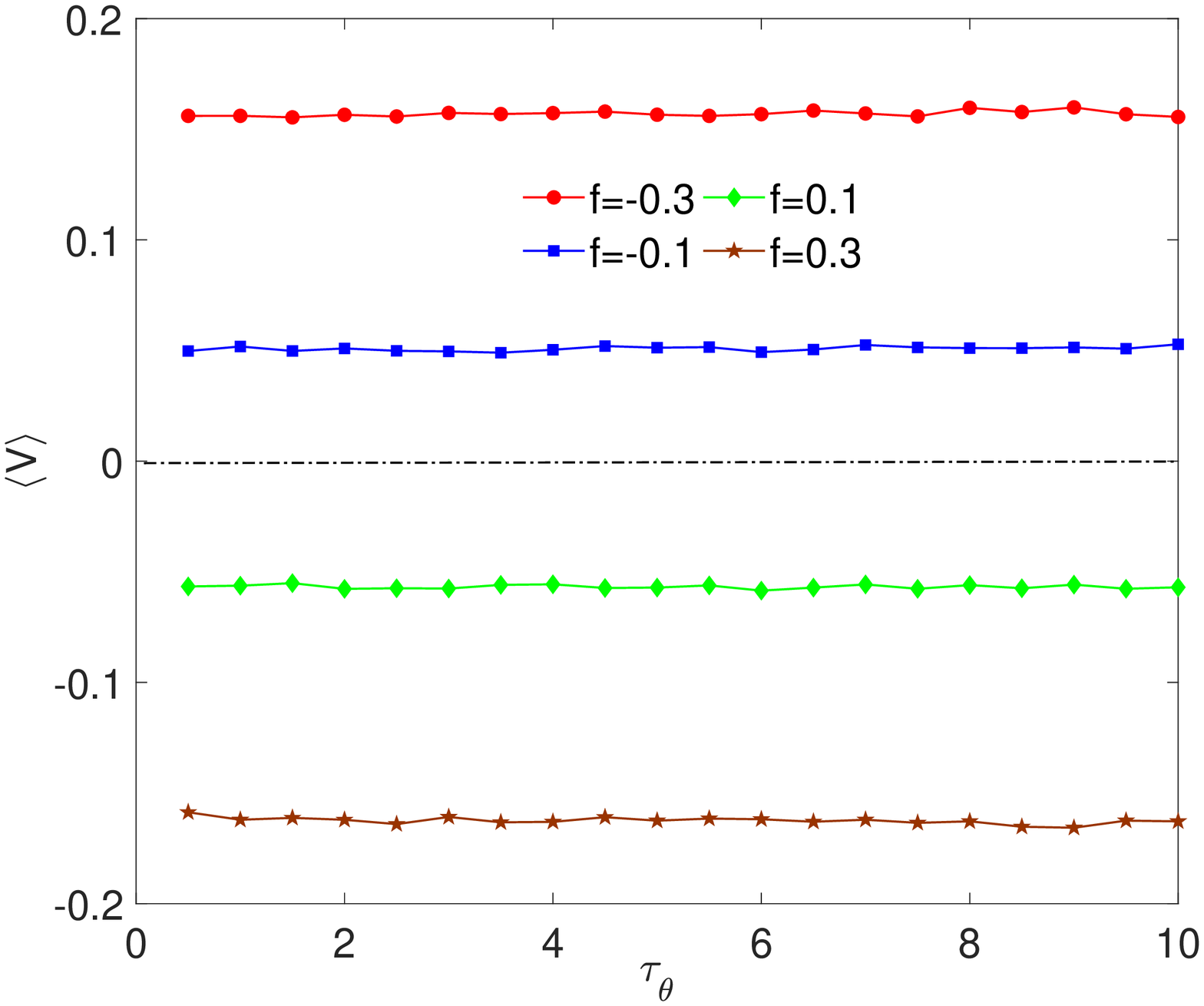}
\caption{The average velocity $\langle V\rangle$ as a function of $\tau_\theta$ for different values of $f$. The other parameters are $\Delta=0.8$, $\varepsilon=0.5$, $\eta=2.0$, $v_0=0.5$, $Q_x=Q_y=Q_\theta=1.0$, $\tau_x=\tau_y=1.0$.}
\label{VTautheta}}
\end{figure}
Figs.\ref{VQtheta} and \ref{VTautheta} display $\langle V\rangle$ as functions of angle noise intensity $Q_\theta$ and self-correlation time $\tau_\theta$ for different $f$, respectively. In these two figures, we find there exits almost no change of $\langle V\rangle$ with increasing $Q_\theta$ and $\tau_\theta$. So the effect of the angle noise is very weak for the transport.

\begin{figure}
\center{
\includegraphics[height=8cm,width=10cm]{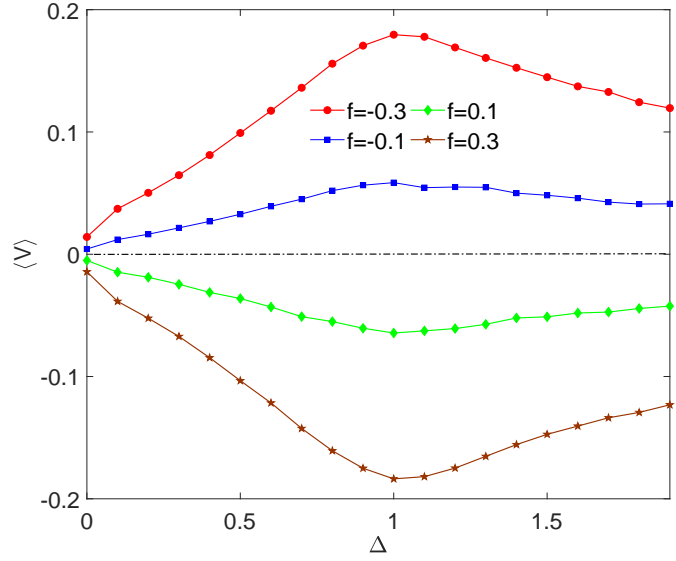}
\caption{The average velocity $\langle V\rangle$ as a function of the size of bottleneck $\Delta$ for different values of $f$.  The other parameters are $\varepsilon=0.5$, $\eta=2.0$, $v_0=0.5$, $Q_x=Q_y=Q_\theta=1.0$, $\tau_x=\tau_y=\tau_\theta=1.0$.}
\label{VDelta}}
\end{figure}
The average velocity $\langle V\rangle$ as a function of the bottleneck(the minimal channel width) $\Delta$ is reported in Fig.\ref{VDelta}. We find $\langle V\rangle\rightarrow0$ when $\Delta\rightarrow0$. This is because the channel becomes many closed cavities when $\Delta\rightarrow0$, and the particle is confined in these cavities.  The directed transport speed $| \langle V\rangle|$ increases with increasing $\Delta$, and reaches a maximum when $\Delta=1$, and then decreases with increasing $\Delta$. So an interesting phenomena appears, that is, large and small size of the bottleneck both restrains the directed transport.

\begin{figure}
\center{
\includegraphics[height=8cm,width=10cm]{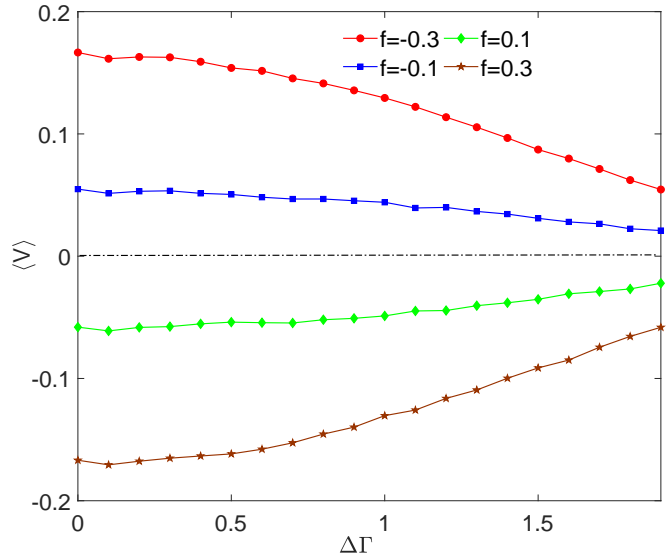}
\caption{The average velocity $\langle V\rangle$ as a function of $\Delta\Gamma$ for different values of $f$. The other parameters are $\Delta=0.8$, $\varepsilon=0.5$, $\eta=2.0$, $v_0=0.5$, $Q_x=Q_y=Q_\theta=1.0$, $\tau_x=\tau_y=\tau_\theta=1.0$.}
\label{VDeltaGamma}}
\end{figure}
The average velocity $\langle V\rangle$ as a function of the asymmetry of the body $\Delta\Gamma$ for different $f$ is reported in Fig.\ref{VDeltaGamma}. We know $\Delta\Gamma$ characterizes the asymmetry of the particle, and the particle is a perfect sphere when $\Delta\Gamma=0$, and the particle is a very needlelike ellipsoid when $\Delta \Gamma\rightarrow\bar\Gamma$. In this figure, we find the average directed transport speed $| \langle V\rangle|$ decreases with increasing $\Delta\Gamma$. So perfect sphere particle is more easier for directed transport than needlelike ellipsoid particle. This result coincides with the common sense.

\begin{figure}
\center{
\includegraphics[height=8cm,width=10cm]{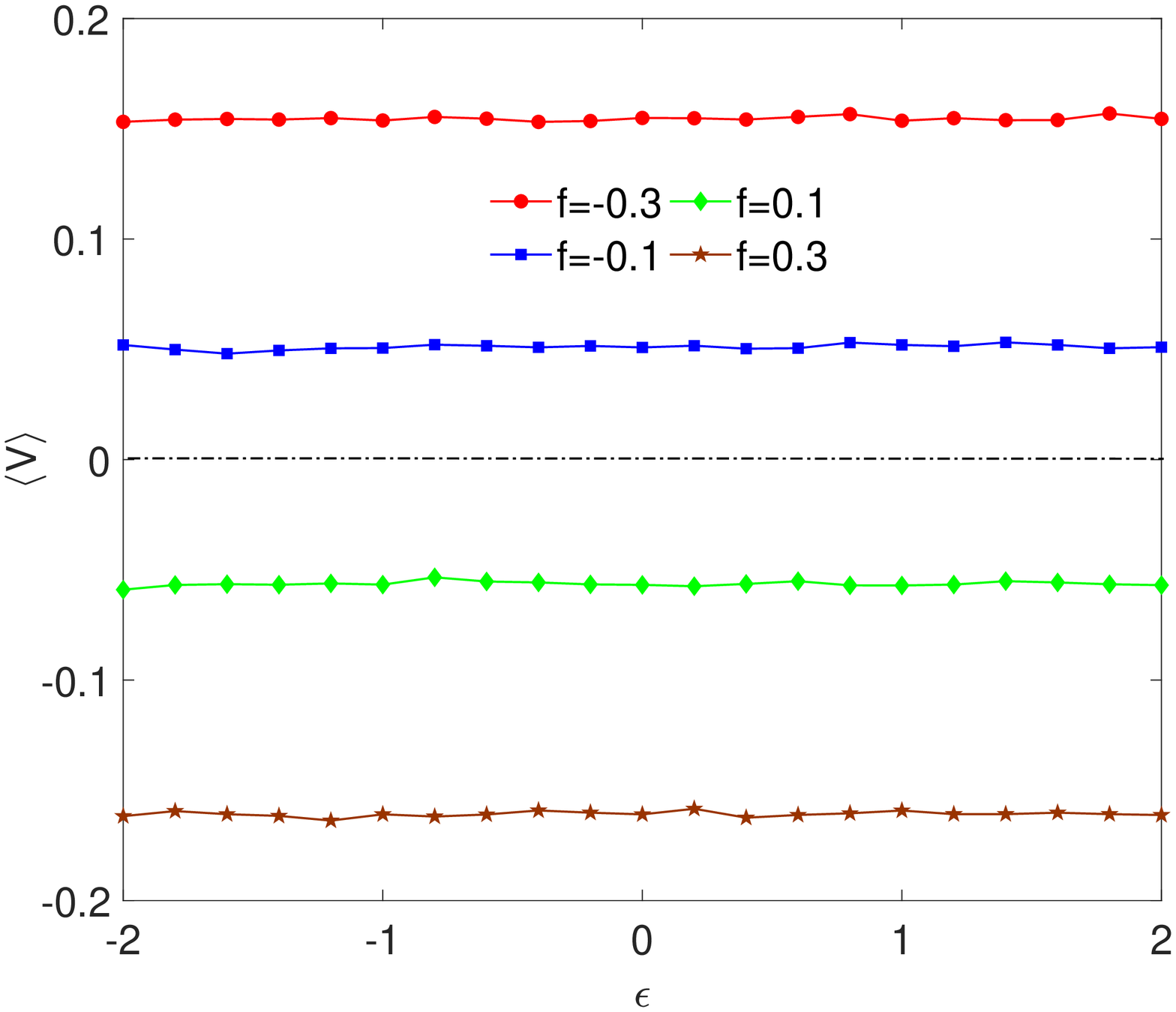}
\caption{The average velocity $\langle V\rangle$ as a function of $\varepsilon$ for different values of $f$.  The other parameters are $\Delta=0.8$, $\eta=2.0$, $v_0=0.5$, $Q_x=Q_y=Q_\theta=1.0$, $\tau_x=\tau_y=\tau_\theta=1.0$.}
\label{VEpsilon}}
\end{figure}
The average velocity $\langle V\rangle$ as a function of the asymmetric parameter $\varepsilon$ for different $f$ is reported in Fig.\ref{VEpsilon}. We find the $\langle V\rangle-\varepsilon$  curve is almost horizontal, this means that the effect of $\varepsilon$ is very weak on the directed transport phenomenon.

\begin{figure}
\center{
\includegraphics[height=8cm,width=10cm]{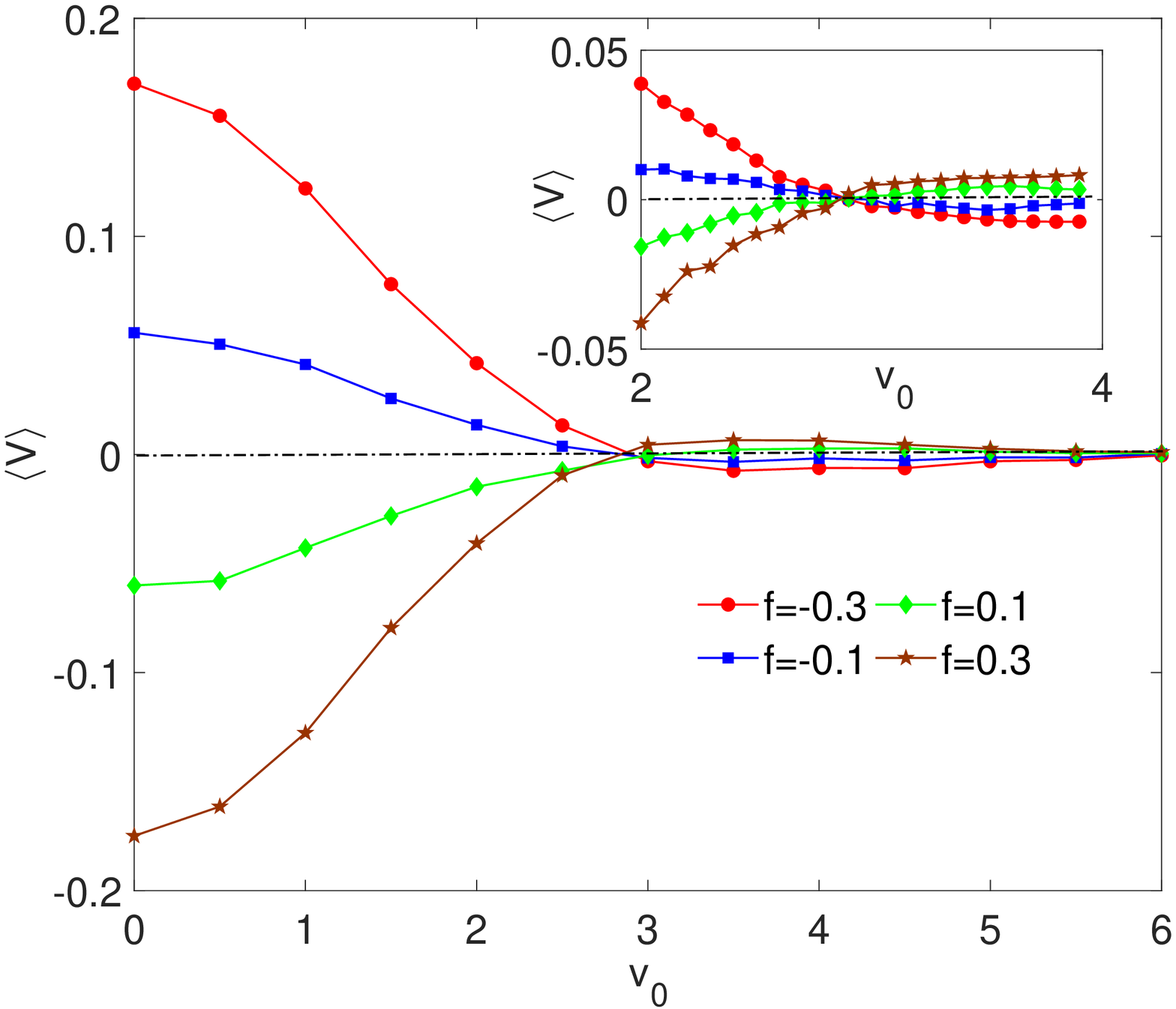}
\caption{The average velocity $\langle V\rangle$ as a function of self-propelled speed $v_0$ for different values of $f$. The other parameters are $\Delta=0.8$, $\varepsilon=0.5$, $\eta=2.0$, $Q_x=Q_y=Q_\theta=1.0$, $\tau_x=\tau_y=\tau_\theta=1.0$.}
\label{Vv0}}
\end{figure}
The average velocity $\langle V\rangle$ as a function of self-propelled speed $v_0$ with different $f$ is reported in Fig.\ref{Vv0}. When the load is negative($f=-0.3$ and $f=-0.1$), inert particle(self-propelled speed $v_0=0$) moves in $+x$ direction($\langle V\rangle>0$), and $\langle V\rangle$ decreases with increasing $v_0$, then the particles changes to moving in $-x$ direction when $2.8<v_0<6.0$, and $\langle V\rangle\rightarrow0$ when $v_0=6.0$. So the transport reverse phenomenon appears with increasing $v_0$. When the load is positive($f=0.1$ and $f=0.3$), inert particle($v_0=0$) moves in $-x$ direction($\langle V\rangle<0$), and the transport reverse phenomenon appears too with increasing $v_0$, and $\langle V\rangle\rightarrow0$ when $v_0=6.0$. In this figure, we can also find negative load leads to directed transport in $+x$ direction when $v_0<3$, but positive load leads to directed transport in $-x$ direction when $v_0<3$. So self-propelled speed $v_0$ can also influence the moving direction of the particles.

\begin{figure}
\center{
\includegraphics[height=8cm,width=10cm]{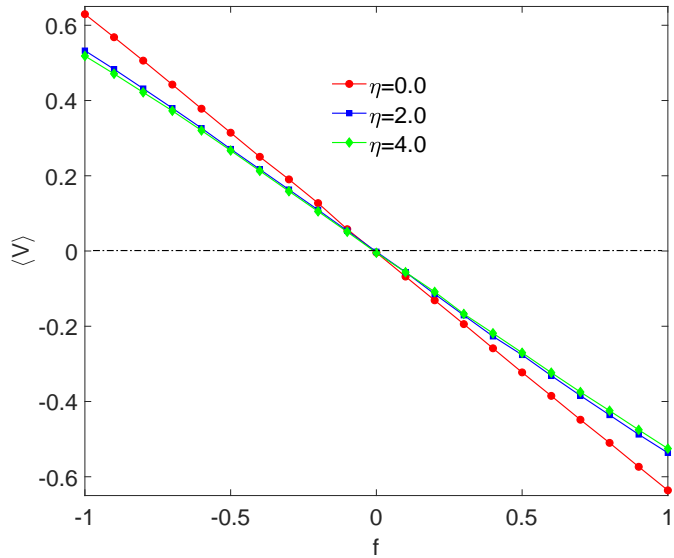}
\caption{The average velocity $\langle V\rangle$ as a function of load $f$ for different $\eta$. The other parameters are $\Delta=0.8$, $\varepsilon=0.5$, $v_0=0.5$, $Q_x=Q_y=Q_\theta=1.0$, $\tau_x=\tau_y=\tau_\theta=1.0$.}
\label{Vf}}
\end{figure}
The average velocity $\langle V\rangle$ as a function of the load $f$ for different $\eta$ is reported in Fig.\ref{Vf}. We find $\langle V\rangle$ decreases monotonically with increasing load $f$. $\langle V\rangle>0$ when $f<0$, but $\langle V\rangle<0$ when $f>0$. This means negative load leads to positive transport(directed transport in $-x$ direction), but positive load leads to negative transport. Directed transport speed $|\langle V\rangle|$ increases with increasing $|f|$, so large load $|f|$ is good for directed transport(in $x$ direction or in $-x$ direction). We know the potential $U$ is inversely proportional to the load $f$(Fig.\ref{Uxy}), so the particles move to the place where potential is lower. In this figure, we can also find the smaller $\eta$, the larger $|\langle V\rangle|$ is.

\section{\label{label4}Conclusions}
In this paper, we numerically investigated the directed transport of self-propelled ellipsoidal particles confined in a smooth channel with potential and colored noise. We find the moving direction is closely linked to the direction of the load when the self-propelled speed is small($v_0<3$). Negative load leads to directed transport in $x$ direction, but positive load leads to directed transport in $-x$ direction. Small and large size of bottleneck restrains the directed transport of the ellipsoidal particle. Large $x$ axis noise intensity inhibits the directed transport. But proper value of $y$ axis noise intensity is good for this phenomenon. The transport reverse phenomenon appears with increasing self-propelled speed $v_0$. The effects of angle noise on the system is negligible.

\section*{Acknowledgments}
Project supported by Natural Science Foundation of Anhui Province(Grant No:1408085QA11).

\end{document}